\documentclass[aps,pra,showpacs,preprintnumbers,onecolumn, amsmath, amssymb, floatfix, notitlepage, 12pt]{revtex4-1}

\usepackage{graphicx}
\usepackage{dcolumn}
\usepackage{bm}
\usepackage{array} 

\usepackage{pdfsync} 
\usepackage{multirow}

\usepackage{enumitem}

\def\app#1#2{%
  \mathrel{%
    \setbox0=\hbox{$#1\sim$}%
    \setbox2=\hbox{%
      \rlap{\hbox{$#1<$}}%
      \lower1.1\ht0\box0%
    }%
    \raise0.25\ht2\box2%
  }%
}

\def\app#1#2{%
  \mathrel{%
    \setbox0=\hbox{$#1\sim$}%
    \setbox2=\hbox{%
      \rlap{\hbox{$#1>$}}%
      \lower1.1\ht0\box0%
    }%
    \raise0.25\ht2\box2%
  }%
}

\setlist{nolistsep} 

\begin{document}


\title{Optimizing the efficiency  of evaporative cooling in optical dipole traps}

\author{Abraham J. Olson}
 \email{olsonaj@purdue.edu}
 \author{Robert J. Niffenegger}%
\author{Yong P. Chen}%
\email{yongchen@purdue.edu}
\affiliation{%
Department of Physics, \\Purdue University, West Lafayette IN 47907
}%

\date{\today}

\begin{abstract}
We present a combined computational and experimental study to optimize the efficiency of evaporative cooling for atoms in optical dipole traps. By employing a kinetic model of evaporation, we provide a strategy for determining the optimal relation between atom temperature, trap depth, and average trap frequency during evaporation given experimental initial conditions. We then experimentally implement a highly efficient evaporation process in an optical dipole trap, showing excellent agreement between the theory and experiment. This method has allowed the creation of pure Bose-Einstein condensates of $^{87}$Rb with 2$\times 10^4$ atoms starting from only  $5\times 10^5$ atoms initially loaded in the optical dipole trap, achieving an evaporation efficiency, $\gamma_{eff}$, of 4.0 during evaporation.

\end{abstract}

\pacs{64.70.fm, 67.85.Hj}
\maketitle

\section{Introduction}

Over the last 30 years, the control and study of ultracold atoms has continued to impress the research community. Initial studies in both optical \cite{Chu_PRL_1986} and magnetic  \cite{Lovelace_Nature_1985} confinement of neutral atoms led to the realization that forced evaporative cooling could create colder atom temperatures \cite{Masuhara_PRL_1988}. Forced evaporation proceeds by a lowering of the trap depth, which allows the hotter atoms to escape the trap while the colder atoms remain and rethermalize. This method of evaporative cooling played a critical role in realizing atomic Bose-Einstein condensates (BECs) \cite{Anderson_Science_1995, Davis_PRL_1995, Bradley_PRL_1995, *Bradley_PRL_1997} and degenerate Fermi gases \cite{DeMarco_Science_1999}. Though primarily used as a tool in the creation of quantum degenerate gases, evaporative cooling also presents interesting physics in itself \cite{Ketterle_AdvAMO_1996, Luiten_PRA_1996, Surkov_PRA_1996, Walraven_book_1996, Sackett_PRA_1997, Yamashita_PRA_2003}. 

Most early experiments in evaporative cooling were performed in magnetic traps. Optical dipole trapping and evaporation, however, have been increasingly popular, because optical traps give access to all magnetic spin states, have less stringent vacuum requirements, and allow the use of Feshbach resonances to modify atomic interactions. A key difference between optical and magnetic forced evaporative cooling is the modification of the trap frequency during evaporation. In magnetic traps, since evaporation is carried out by using an rf ``knife'' to remove atoms from the trap, the trap frequency can remain constant as the trap depth is lowered \cite{Davis_PRL_1995a}. In optical dipole traps, the simplest forced evaporative cooling is done by lowering the trapping laser power. In contrast to magnetic trapping, for such evaporation the average trap frequency, $\bar{\omega}$, is reduced as $\bar{\omega} \propto U^{0.5}$ where $U$ is the trap depth \cite{Barrett_PRL_2001, OHara_PRA_2001}. Since the elastic collision rate (and thus the evaporation rate) decreases with decreasing $\bar{\omega}$, efficient all-optical evaporative cooling can be hindered by stagnation of the cooling process. To address this limitation, various techniques have been developed to change the relationship between the trap frequency and trap depth during evaporation in optical dipole traps, such as the zoom lens trap \cite{Kinoshita_PRA_2005}, the tilted trap \cite{Hung_PRA_2008,Clement_PRA_2009}, and transitioning from a single beam to cross-beam optical dipole trap geometry \cite{Arnold_OptComm_2011}.

The typical goal of evaporative cooling is to increase the number of atoms that reach quantum degeneracy. Atom losses via three-body recombination and one-body background collisions limit the attainable efficiency for most experimental evaporative cooling efforts. To optimize evaporation, the rates of three-body and one-body losses should be kept small compared to the rate of evaporation. Generally, if evaporation proceeds too slowly, one-body losses become the major loss mechanism and limit achievable efficiency. If evaporation proceeds rapidly with nearly constant trapping frequency, three-body losses may be dominant. 

In this paper, we report a general strategy for optimizing the efficiency of evaporation in optical dipole traps. This strategy involves selecting a relationship between the trap frequency, trap depth, and atom cloud temperature during evaporation to minimize one-body and three-body losses, and thereby maximize the number of atoms that reach the desired final phase space density. We first introduce a theoretical model for evaporative cooling. We demonstrate using this model to find optimal evaporative cooling routes for experiments. We then present results utilizing this strategy in our experiment, in which a highly efficient optical evaporative cooling is achieved. 

A few definitions will aid in the ensuing discussion. We parametrize the relationship between the weakening of the trap frequency and the trap depth by $\nu$, where $\bar{\omega} \propto U^{\nu}$. The ratio of trap depth, $U$, to atom cloud temperature, $T$, is given by $\eta = U/ k_B T$ where $k_B$ is the Boltzmann's constant. Optimal evaporation is achieved by selecting these two parameters, $\eta$ and $\nu$, for the evaporation route such that rates of one-body and three-body losses, $\Gamma_{1B}$ and $\Gamma_{3B}$, are kept small. The efficiency of evaporation is quantified by $\gamma_{eff} = -\ln (\rho_f/\rho_i) / \ln (N_f/N_i)$, where $\rho = n_0 \lambda_{dB}^3$ is the phase space density, $N$ is the number of atoms in the trap, $n_0= N\bar{\omega}^3 [m/ (2\pi k_B T)]^{3/2}$ is the calculated peak atomic density, $\lambda_{dB} = \sqrt{2\pi \hbar^2 / m k_B T }$ is the thermal deBroglie wavelength, $\bar{\omega}$ is the geometric mean of the trap frequencies, $m$ is the atomic mass, and $\hbar$ is the reduced Planck's constant. Many previous experiments typically achieve values of $\gamma_{eff} \approx 2.5-3.5$. Our optimized scheme realizes $\gamma_{eff} =4.0$, and our theoretical model gives a guide for other experiments to optimize their efficiencies.

\section{Theory}
\label{sec:theory}
We first present the theory employed to model evaporative cooling of ultracold atomic gases. While scaling laws have been derived to describe evaporative cooling in optical dipole traps and are helpful in gaining a qualitative understanding, the analyses are limited by their neglect of losses or by only treating specific cases (i.e. $\nu = 0.5$) \cite{OHara_PRA_2001, Hung_PRA_2008}. To develop our strategy of optimizing the evaporative cooling, we employ a kinetic theory of evaporative cooling \cite{Luiten_PRA_1996, Pinkse_PRA_1998, Yamashita_PRA_2003,Yamashita_LP_2003}. In the kinetic theory approach, a truncated Boltzmann distribution is used to describe the distribution of atoms in the trapping potential, $U(\vec{r})$: 

\begin{equation}\label{eqn:fdist}
f(\vec{r}, \vec{p}) = n_0 \lambda_{dB}^3 \exp \left[ -\left( U(\vec{r}) + p^2/2m\right) /k_B T \right] \Theta \left( \eta k_B T - U(\vec{r}) - p^2 / 2m \right)
\end{equation}
In the deep trap limit ($\eta > 6$ for three-dimensional (3D) harmonic traps) the truncation effects are small \cite{Luiten_PRA_1996} and the Heaviside step function, $\Theta$, in the distribution can be replaced by unity for calculations of the atom number and energy density. 

With this simplification, the spatial density is
\begin{equation}\label{eqn:ndist}
n(\vec{r}) = \frac{1}{(2 \pi \hbar)^3} \int f(\vec{r}, \vec{p}) d^3\vec{p} = n_0 \exp \left( -U(\vec{r}) / k_B T \right)
\end{equation}
and the energy density of the atoms in the trap (neglecting interactions) is
\begin{equation}\label{eqn:edist}
\begin{split}
e(\vec{r}) &= \frac{1}{(2 \pi \hbar)^3} \int \left( \frac{p^2}{2 m} + U(\vec{r}) \right) f(\vec{r}, \vec{p}) d^3\vec{p} \\
 & = \frac{3}{2} n_0 k_B T \exp \left( -U(\vec{r})/ k_B T \right) + U(\vec{r}) n(\vec{r})
\end{split}
\end{equation}
The total energy in this deep trap limit is $E = 3 N k_B T$, and the atom number and energy evolution during evaporation can be modeled by 
\begin{subequations}
\label{eqn:NandEevolutions}
 \begin{align}
  \dot{N} &= \dot{N}_{ev} + \dot{N}_{\theta}+\dot{N}_{1B} + \dot{N}_{3B} \\
  \dot{E} &= \dot{E}_{ev} + \dot{E}_{\theta}+\dot{E}_{1B} + \dot{E}_{3B}
 \end{align}
\end{subequations}
where, on the right-hand side, the first term in each equation accounts for effects of evaporation, the second for trap shape changes (denoted by $\theta$), and the final two terms for one-body loss due to background collisions and three-body loss, respectively. Effects of dipolar loss, heating from fluctuations in the trapping potential, and off-resonant photon scattering are small for the experiments considered here and thus neglected. 

Using the kinetic theory with Eqs. \ref{eqn:ndist} and \ref{eqn:edist}, we obtain specific expressions for each term in Eq. \ref{eqn:NandEevolutions}. For our theory, we use a 3D harmonic trap approximation, and express the potential $U(\vec{r}) = \frac{1}{2} m \left( \omega_x^2 x^2 + \omega_y^2 y^2 + \omega_z^2 z^2\right)$ and thus the mean trap frequency as $\bar{\omega} = (\omega_x \omega_y \omega_z)^{1/3}$. While our specific trapping configuration does introduce a slightly anharmonic trap shape (see Fig.~\ref{fig:apparatusModel}), we found it to be a small perturbation due to the high trap depth maintained during evaporation and that employing a 3D harmonic trap in the theory provides a good approximation that models well our experimental results.

\subsection{Collision dependent terms}
The truncated Boltzmann distribution treatment of evaporation assumes that any atoms excited to energy greater than the trap depth by elastic collisions are evaporated. The rate of the evaporation for a 3D harmonic trap is well approximated by $\Gamma_{ev} \approx (\eta-4) e^{-\eta} \Gamma_{el}$ when $\eta \geq 6$ \cite{Luiten_PRA_1996, Hung_PRA_2008}. Here $\Gamma_{el}$ is the elastic collision rate of the atoms, and in a 3D harmonic trap $\Gamma_{el}= n \sigma \bar{v}/\left(2 \sqrt{2}\right)$, where $\sigma =8\pi a_s^2$ is the elastic cross section for identical bosons, $a_s$ is the $s$-wave scattering length ($a_s=98 a_0$ for $^{87}$Rb where $a_0$ is the Bohr radius  \cite{VanKempen_PRL_2002}), and $\bar{v} = 4 (k_B T/ \pi m)^{1/2}$ is the average relative velocity of the atoms. Each evaporated atom carries away energy greater than the trap depth, and the average energy removed by each evaporated atom is $(\eta + \kappa ) k_{B} T$, where $\kappa \approx (\eta - 5)/(\eta -4)$ for a 3D harmonic trap in the deep trap limit \cite{Luiten_PRA_1996, Hung_PRA_2008}.

There are no bounds on the achievable efficiency of evaporation for an atomic gas with only elastic collisions. For actual experiments, however, the effects on the atom number and energy due to one-body and three-body losses limit the attainable evaporation efficiency and must be considered. Such effects are expressed as
\begin{equation}\label{eqn:lossesRateN}
\begin{split}
\dot{N}_{1B} + \dot{N}_{3B} & =  - \Gamma_{1B} \int n(\vec{r}) d^3\vec{r} - L_{3B} \int n(\vec{r})^3 d^3\vec{r} \\
 & =-\Gamma_{1B} N - \Gamma_{3B} N 
 \end{split}
\end{equation}
\begin{equation}
\begin{split}
\dot{E}_{1B} + \dot{E}_{3B} & = - \Gamma_{1B} \int e(\vec{r}) d^3\vec{r} - L_{3B} \int n(\vec{r})^2 e(\vec{r}) d^3\vec{r} \\
 & = -\Gamma_{1B} E -\Gamma_{3B} \frac{2}{3} E 
\end{split}
\end{equation}
where $\Gamma_{3B} = L_{3B} n_0^2 /(3 \sqrt{3})$ ($L_{3B} = 4.3(\pm 1.8)\times 10^{-29}$cm$^6$/s for $^{87}$Rb in the $F=1$ ground state \cite{Burt_PRL_1997}). $\Gamma_{1B}$ is typically dominated by the loss rate due to background collisions, and is set by the vacuum conditions of the experimental chamber. It is measured experimentally from the trapped atom loss rate in a very deep trap with low atom density \footnote{The results show the importance of finding the trap averaged values given the dimensionality and trap type used in the experiment to obtain quantitative agreement between the theory and experiments. $\Gamma_{el}$ is sometimes reported as $n \sigma \bar{v}$, but in a 3D harmonic potential is reduced by $2\sqrt{2}$. Similarly, since three-body collisions occur in the denser, less energetic regions of the trap, the average energy loss per atom from three-body collisions is two-thirds of the average energy of an atom in the trap.}. For other trapping potentials types (e.g. linear 1D), similar equations can be obtained to find $\Gamma_{el}$ and $\Gamma_{3B}$ \cite{Luiten_PRA_1996, Pinkse_PRA_1998}. 

\begin{figure}[b!]
  \includegraphics[width=0.5\textwidth]{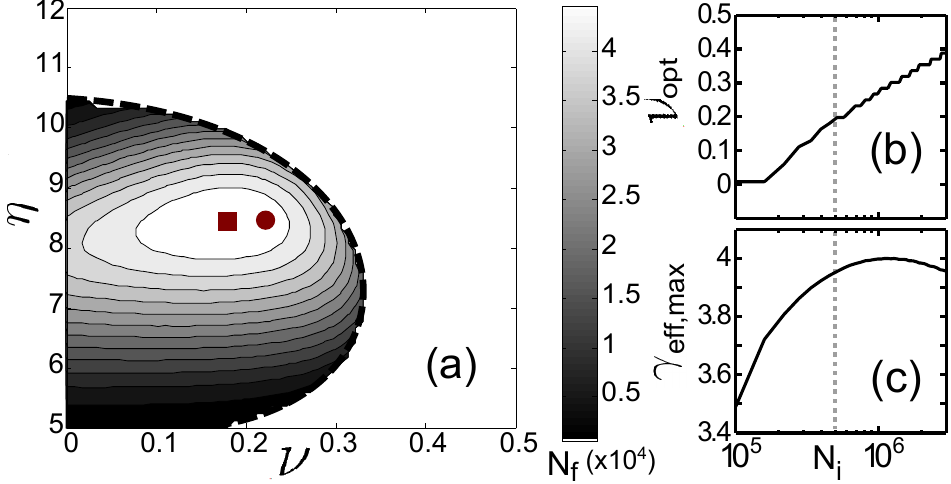}
  \caption{(Color online) (a) Numerical solution of the final number of atoms, $N_f$, to reach a final phase-space density of $\rho_f = 2.6$ as a function of $\eta$ and $\nu$ given our experimental initial conditions.  The total evaporation time is bounded to be $T_{evap}<10$ s (area bounded by the thick, dashed line). The maximum efficiency, $\gamma_{eff,max}$, is 3.93 and is achieved at the maximum in $N_f$ (square). Our experimental implementation employed a slightly different set of values of $\eta$ and $\nu$ (circle) that results in a simulated efficiency of $\gamma_{eff}=3.85$.  Initial conditions of evaporation are $T = 60$ $\mu$K, $N=5\times 10^5$, $\bar{\omega}=2\pi \times 1000$ Hz, $\Gamma_{1B} = 1/12$ sec$^{-1}$. Since $N_i$, $\rho_i$ and $\rho_f$ are fixed, $\gamma_{eff}$ depends only on $N_f$. (b) and (c): Effect of varying the initial atom number, $N_i$, while keeping other initial conditions the same as in (a) on the optimal value of $\nu$ for evaporation, (b), and resulting $\gamma_{eff,max}$, (c). The grey dashed line marks the value of $N_i$ in our experiment.}
	\label{fig:optimalgamma}
\end{figure}

\subsection{Changes in the trap shape}
The changing trap frequencies can cause atom loss by the removal of atoms with energy near the trap depth, so-called ``spilling'' \cite{Pinkse_PRA_1998}. It can also cause adiabatic changes in the energy. For deep traps ($\eta>6$), the occupation of states that are spilled during the evaporation is small, and thus $\dot{N_{\theta}} \approx 0$. The adiabatic work done on the trap, however, is non-negligible. Generally, adiabatic changes in the energy due to changes in the trap characteristics are modeled in the kinetic theory \cite{Pinkse_PRA_1998} by
\begin{equation}\label{eqn:adiabaticE}
\dot{E}_{ad} = -\frac{N k_B T}{N/n_0} \left( \frac{\partial (N/n_0)}{\partial \theta}\right)_{T} \dot{\theta}
\end{equation}
where $\theta$ is some trap parameter. The $N/n_0$ terms in the equation can be understood as the effective volume of the trap. In these experiments, $\bar{\omega}$ is the changing trap parameter. We thus replace $\theta$ with $\bar{\omega}$ in Eq.~\ref{eqn:adiabaticE} to obtain: 
\begin{equation}
\dot{E}_{ad} = 3 N k_B T \frac{\dot{\bar{\omega}}}{\bar{\omega}} = \nu E \frac{\dot{U}}{U}=\nu E \frac{\dot{T}}{T}
\end{equation}
This can be intuitively understood as the work done by the atoms on the trap as the trap frequency is adiabatically weakened, and for a fixed trap frequency ($\nu=0$) this term would be zero. 

\subsection{Combined equations and simulation results}
Combining the terms discussed above, the energy and atom number evolution equations for evaporative cooling in deep, 3D, harmonic traps take the form:
\begin{equation} \label{Eqn:energy}
\dot{E} =  -N \Gamma_{ev} (\eta+\kappa) k_B T + \nu E \frac{\dot{T}}{T} -\Gamma_{1B} E -\Gamma_{3B} \frac{2}{3} E
\end{equation}
\begin{equation}\label{Eqn:number}
\dot{N} = -( \Gamma_{ev} + \Gamma_{1B} + \Gamma_{3B})N
\end{equation} 
By numerically solving this model given the initial experimental conditions after loading atoms in the optical dipole trap, the ideal values of $\eta$ and $\nu$ can be found to maximize the number of atoms that reach quantum degeneracy. We note that in this work we assume constant $\eta$ and $\nu$ for optimizing the evaporation process. At the cost of additional complexity, $\eta$ and $\nu$ could be varied during evaporation to achieve slightly more efficient evaporation \footnote{For the initial conditions of our experiment, we find that implementing a variable $\nu (t)$ during evaporation gives only a small gain ($\approx 0.1$) in the numerically computed $\gamma_{eff}$.}.

Figure~\ref{fig:optimalgamma}a shows a calculation given our initial conditions and evaporating to $\rho_f = 2.6$ for different values of $\eta$ and $\nu$. We find a region of optimal $\eta$ and $\nu$ centered around $\eta=8.4$, $\nu=0.18$. As a further study, in Figs.~\ref{fig:optimalgamma}(b) and \ref{fig:optimalgamma}(c) we find that a different initial number, $N_i$, of atoms loaded in the optical dipole trap changes the optimal $\nu$ as well as the maximum $\gamma_{eff}$ (while the optimal $\eta$ is minimally affected by $N_i$) \footnote{Values of $\nu < 0$ are theoretically possible to treat, but are not considered here as experimental limitations in available laser power usually restrict $\nu$ to greater than $0$.}.

\section{Experiment}
\begin{figure}[htb]
  \includegraphics[width=0.45\textwidth]{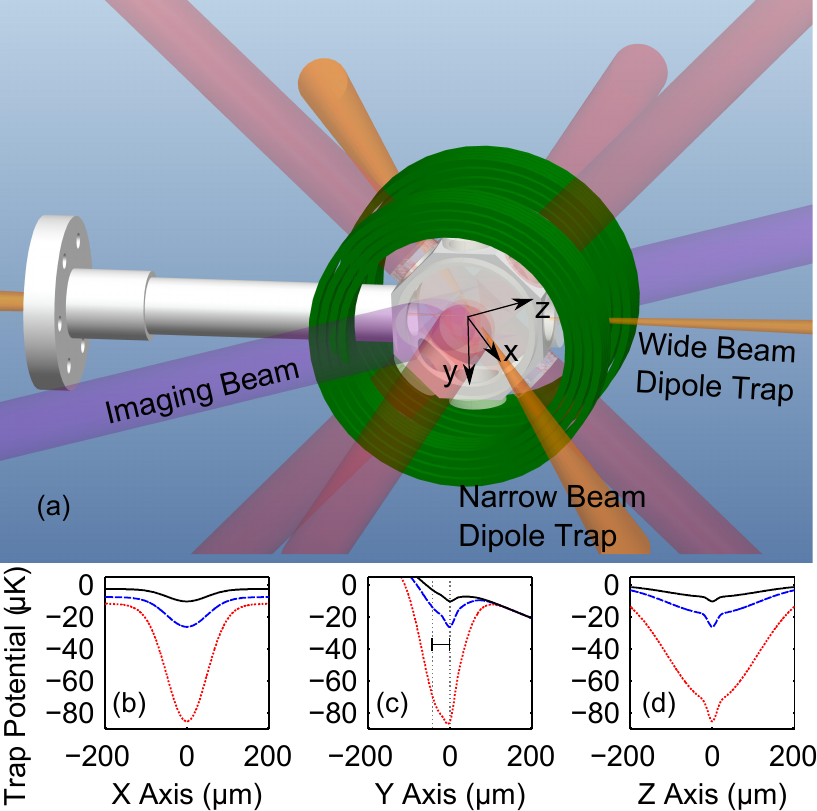}
  \caption{(Color online) A schematic of the experimental setup (a), and calculated trapping potentials for various optical dipole trap laser powers along three axes (b-d). For (a), the magneto-optical trap magnetic field coils are in green, dipole trapping beams in orange, MOT trapping beams in red, and imaging beam in purple. Not pictured are the bias coils, which are used to cancel the earth's magnetic field and apply constant offset magnetic fields. Gravity is along the positive y-axis. The calculated trapping potentials indicated by the dotted red, dashed blue, and solid black lines are shown in (b-d) for various combinations of the optical trap beams' powers, corresponding to $(P_{wide},P_{narrow}) =$(12W, 0.20W), (3W, 0.09W), (1.25W, 0.058W), respectively. The 45 $\mu$m offset between the wide and narrow beams in the $y$direction, marked by the two vertical dashed lines, can be seen in (c). }
	\label{fig:apparatusModel}
\end{figure}

We experimentally implement this evaporative cooling scheme using a misaligned crossed-beam optical dipole force trap (MACRO-FORT \cite{Clement_PRA_2009}), which allows precise tailoring of $\nu$. As our laser source for the optical dipole trap, we use a single-frequency, single spatial mode, erbium fiber laser with wavelength at 1550 nm (IPG Photonics ELR-50-1550-LP-SF). The wide dipole trap beam has a beam waist of 88~$\mu$m and 18~W of initial power ($P_{wide}$). The narrow beam (used to produce the ``dimple'' potential) has a waist of 20 $\mu$m, 9 W of initial power ($P_{narrow}$), crosses the wide beam at an angle of 66$^{\circ}$, and is offset 45~$\mu$m radially from the wide beam's focus (see Fig.~\ref{fig:apparatusModel}). 

For this optical dipole trap geometry, $\nu$ is determined by both the offset of the narrow beam from the wide beam and by the relative powers of each beam. Therefore, achieving different values of $\nu$ for an evaporative route is simply changed by adjusting the laser powers or the offset of the beams. Limitations in the laser power available restrict $\nu$ to be between 0.15 and 0.50. Different values of $\eta$ are simply achieved by varying the total time of the evaporation route. The moderate tolerance of the parameters as seen in Fig.~\ref{fig:optimalgamma} allow for some flexibility in the experimental choice of $\eta$ and $\nu$, and we have designed our experimental evaporation ramp to have $\eta \approx 8.5$ and $\nu=0.22$, so as to preferentially avoid three-body losses while still maintaining a high efficiency.

\begin{figure}[htb]
  \includegraphics[width=0.45\textwidth]{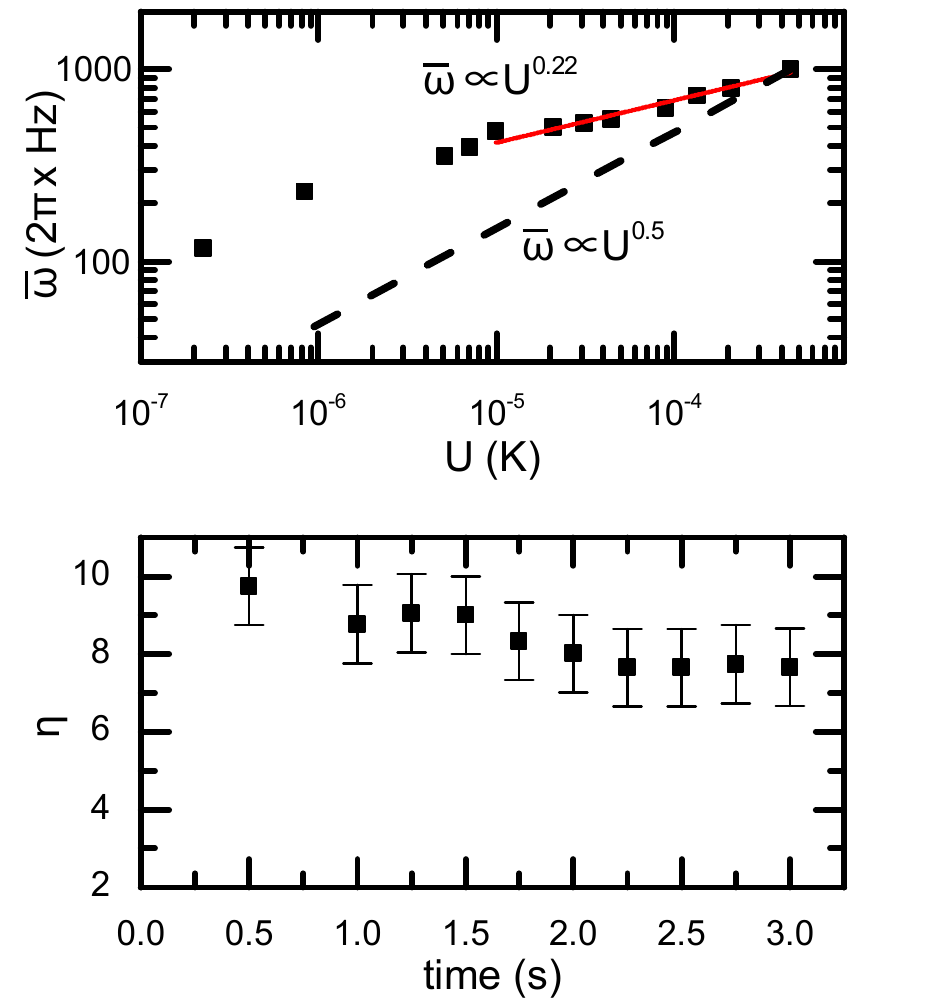}
  \caption{(Color online) (a) Measured average trap frequency, $\bar{\omega}$, vs trap depth, $U$, for the near-optimal evaporation trajectory used in our experiment. The trap frequencies were measured by the parametric heating method. The experiment used $\nu=0.22$ for the first 3 seconds of evaporation. For comparison, the dashed line indicates the $\bar{\omega}\propto U^{0.5}$ relationship used in conventional optical evaporative cooling. (b) Extracted $\eta$ during evaporation based on the $T$ of atoms measured from time-of-flight, and trap depth $U$ calculated (assuming ideal Gaussian beams) from the measured dipole trap laser powers and beam geometry.}
	\label{fig:TrapFreqVsU}
\end{figure}

\begin{figure}[htb]
  \includegraphics[width=0.4\textwidth]{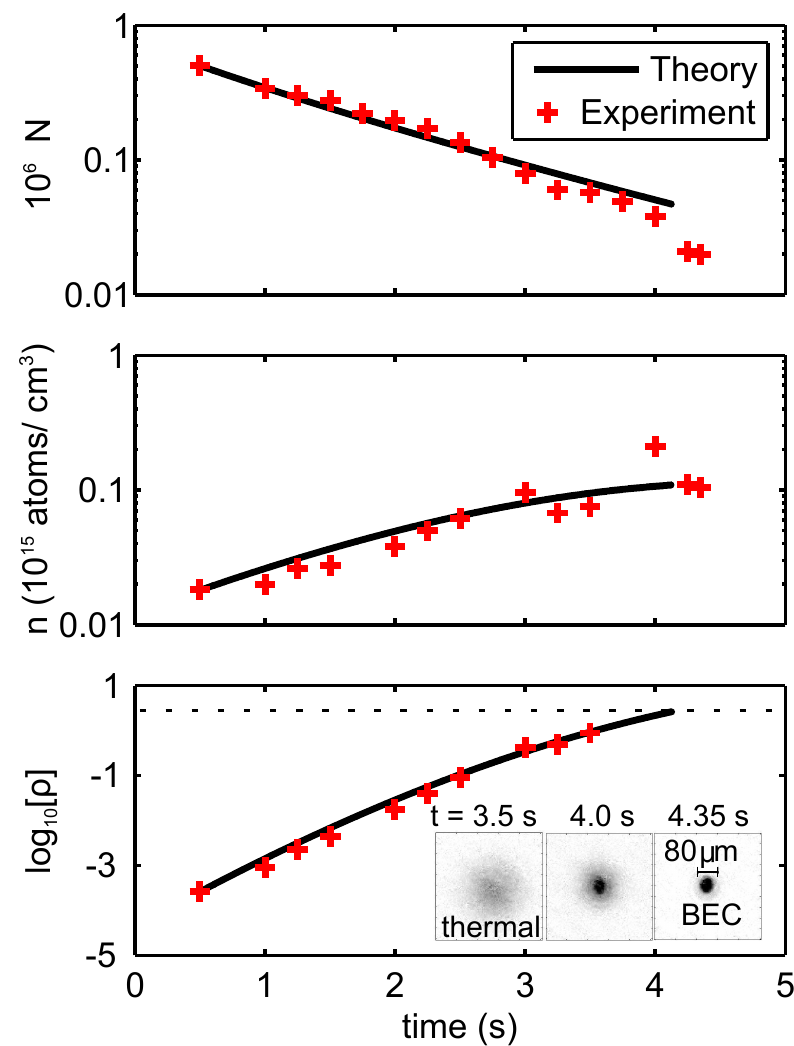}
  \caption{(Color online) The experimentally measured atom number (a), atomic density (b), and  phase-space density (c), vs time during evaporation. The theoretical calculations utilize the model of Eqs.~\ref{Eqn:energy} and \ref{Eqn:number} for $\nu=0.22$, $\eta = 8.5$, $\bar{\omega_i}=2\pi \times 1000$ Hz, and assuming our experimental initial conditions. The slight overestimate of the theoretical $n$ and $\rho$ during the first 2.5 seconds is likely due to the presence of atoms in the ``wings'' of the narrow beam trap (e.g. see Ref. \cite{Arnold_OptComm_2011}). The insets in (c) show resonant absorption images (taken after 10 ms of time-of-flight expansion) of the atomic cloud at three different times along the evaporation ramp, showing the thermal cloud to BEC transition. }
	\label{fig:experimentalResults}
\end{figure}

To confirm that we follow the optimal evaporation route modeled in Fig.~\ref{fig:optimalgamma}, measurements of the trap frequency, atom temperature, and a calculation of the trap depth allow us to determine the experimental $\eta$ and $\nu$ values during the evaporation ramp (Fig.~\ref{fig:TrapFreqVsU}). After loading the optical dipole trap with $5\times 10^5$ atoms as described in the Appendix, forced evaporation with $\nu=0.22$ and $\eta \approx 8-10$ (noting the relative insensitivity of $\gamma_{eff}$ to small deviations of $\eta$ and $\nu$ from the optimal target values in Fig.~\ref{fig:optimalgamma}) was implemented by exponentially decreasing the power in both beams over the course of 3 seconds to 12~W and 0.068~W respectively.  At this point, there are $10^5$ atoms at just over 1~$\mu$K (see Fig.~\ref{fig:experimentalResults}). The average trap frequency is 475~Hz so that the BEC critical temperature is $T_C \approx 900 \text{ nK}$ (where $k_B T_c = \hbar \bar{\omega}( N/ 1.202)^{1/3}$). Continuing to evaporate in such a tight trap, however, would yield high three-body losses. Thus, over the final 1.35~seconds, the power in the wide beam is lowered to 1.25~W and the narrow beam to 0.020~W. This results in a nearly pure BEC of $2\times 10^4$ atoms in a trap with measured frequencies of $2 \pi \times$(60, 100, 270) Hz. During the transition to BEC, the evaporative process is aided by bosonic enhancement \cite{Miesner_Science_1998,Kohl_PRL_2002} and no longer well described by the classical evaporative theory as described in Section~\ref{sec:theory}. This yields efficiencies even higher than the simulated values.

\begin{figure}[htbp]
  \includegraphics[width=0.45\textwidth]{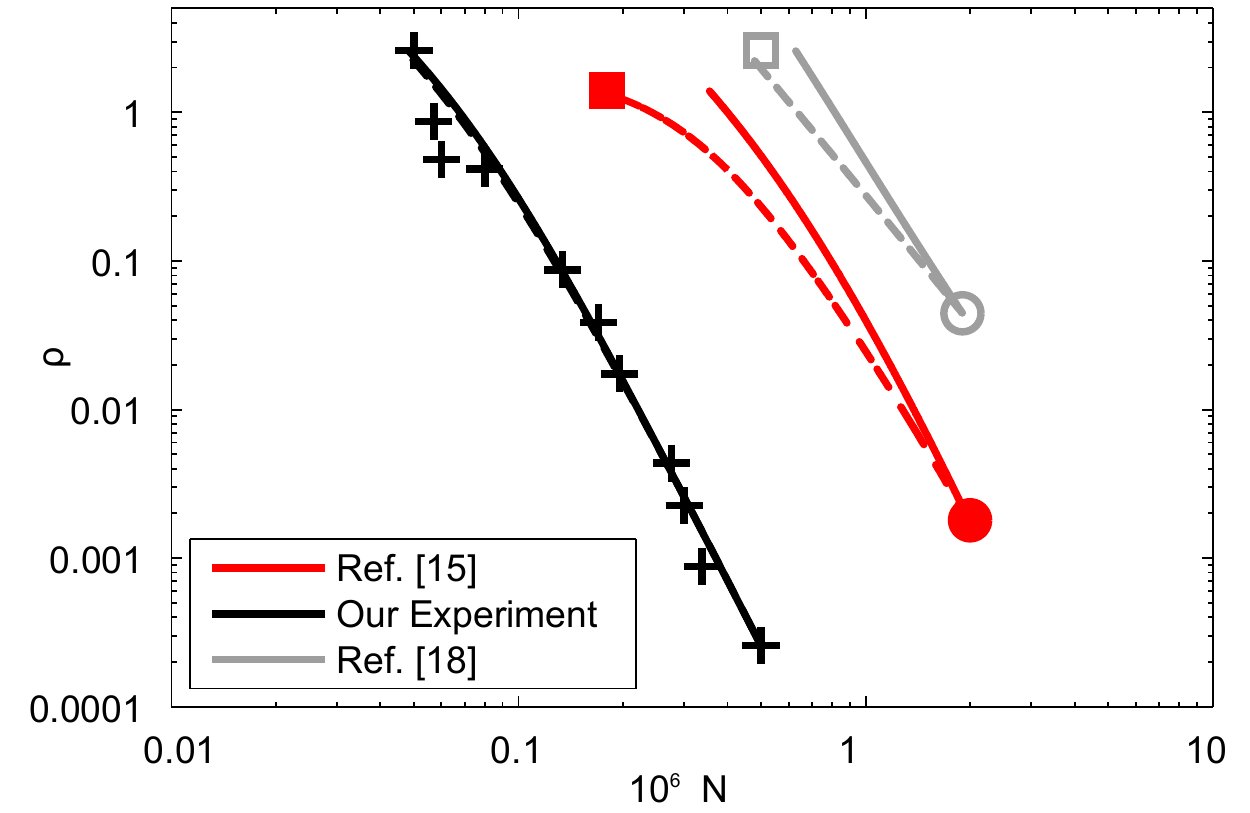}
  \caption{(Color online) Comparison between experimental and simulated evaporation trajectories for three different experiments. The circles (squares) indicate the experimental results of the initial (final) atom number and phase-space density from Refs.~\cite{Barrett_PRL_2001} (red, filled shapes)  and \cite{Hung_PRA_2008} (light grey, open shapes).  Dashed lines indicate simulation of evaporation given reported experimental initial conditions and evaporation parameters (see Table \ref{tab:evapCompare}). Solid lines indicate the simulated optimized evaporation route by using the optimizing strategy for $\eta$ and $\nu$ developed in this paper. For our work, the solid line overlaps the dashed line almost completely. The results of the simulation match each of the experiments well, and further details of the quantitative agreement are presented in Table \ref{tab:evapCompare}.}
	\label{fig:compareExp}
\end{figure}

The strategy of optimizing the evaporative cooling by selecting the optimum values of $\eta$ and $\nu$ is general and not limited to the specific trap configuration used in our work. The theory we have presented above is limited to deep traps that are approximately 3D harmonic. Many optical dipole trap implementations, however, are well approximated by these assumptions, because tighter confinement afforded by optical dipole traps allow for rapid evaporation even with $\eta>6$, and the intensity profiles of the laser beams are harmonic to first order. To demonstrate the general applicability of the theory and strategy, we model two other experiments of evaporation in optical dipole traps, one in $^{133}$Cs by Hung \emph{et al.} \cite{Hung_PRA_2008} and the other in $^{87}$Rb by Barrett \emph{et al.} \cite{Barrett_PRL_2001} (see Fig.~\ref{fig:compareExp}). We find excellent agreement between our theoretical simulations and their experimental results, and our simulation results show that both experiments could further improve their evaporation efficiency by optimizing both $\eta$ and $\nu$ \footnote{The total time of evaporation is important when considering the cycle time of an apparatus which produces BECs. The optimized schemes presented here take $0.7$ and $10$ seconds for the Barrett \emph{et al.} and Hung \emph{et al.} experiments, respectively.}. 

In summary, we have described a scheme to optimize the efficiency of evaporative cooling by optimizing the relation between the trap frequency, trap depth, and atom temperature during the evaporation process. As a result, we achieve a $\gamma_{eff}=4.0$, a high value for all-optical evaporation. Different from previous treatments of evaporative cooling in optical dipole traps \cite{OHara_PRA_2001, Hung_PRA_2008, Clement_PRA_2009}, this approach includes atom losses in modeling the evaporative process and optimize both $\eta$ and $\nu$ for our experimental conditions. Different from the evaporation optimization scheme in Ref.~\cite{Yamashita_PRA_2003}, our scheme optimizes evaporation by tuning $\nu$ and $\eta$ for the whole evaporation ramp, rather than just changing $\eta$ during evaporation. In addition, we have highlighted the utility of experimental techniques (such as MACRO-FORTs) that allow selecting the $\eta$ and $\nu$ values for all-optical evaporation, and we demonstrate that BECs can be obtained even when starting from a small number ($5 \times 10^5$) of trapped atoms. The method shown here of optimizing the efficiency by finding optimal $\eta$ and $\nu$ is not specific to $^{87}$Rb, and can be applied to achieve optimal $\gamma_{eff}$ for any atomic species that is evaporatively cooled in optical traps. 

\begingroup
\squeezetable
\begin{table*}[htbp]
\begin{tabular}{*{12}{c}}
  \hline
  Ref & $T_i$ ($\mu$K) & $N_i$ & $\omega_i$ (Hz) & $\Gamma_{1B}$ (s$^{-1}$) & $\rho_i$ & $\rho_f$ & ~ & $\eta$ & $\nu$ & $N_f$ & $\gamma_{eff}$ \\
  \hline
  \textbf{Ref. \cite{Barrett_PRL_2001}} & $75$  & $2{\times} 10^6$ & $2 \pi {\times} 1500$ & $1/6$ & 0.0018 & 1.4 & $\left\{\begin{tabular}{@{\ }l@{}}
    \emph{exp} \\ \emph{sim} \\ \emph{opt} \end{tabular}\right.$ & \begin{tabular}{@{\ }l@{}}
    -- \\ 7.2$^{*}$ \\ $8.3$ \end{tabular} & \begin{tabular}{@{\ }l@{}}
    0.50 \\ 0.50 \\ $0.39$ \end{tabular} & \begin{tabular}{@{\ }l@{}}
    $1.8 {\times} 10^5$ \\ $1.7 {\times} 10^5$  \\ $3.6 {\times} 10^5$ \end{tabular} & \begin{tabular}{@{\ }l@{}}
    2.76 \\ 2.67 \\ 3.85 \end{tabular} \\
    
      \textbf{Ref. \cite{Hung_PRA_2008}} & $0.470$  & $1.9 {\times} 10^6$ & $2 \pi {\times} (17{\cdot}34{\cdot}38)^{1/3}$ & $1/50$ & $0.0445$ & 2.6 & $\left\{\begin{tabular}{@{\ }l@{}}
    \emph{exp} \\ \emph{sim} \\ \emph{opt} \end{tabular}\right.$ & \begin{tabular}{@{\ }l@{}}
    6.5 \\ 6.5 \\ 8.2 \end{tabular} & \begin{tabular}{@{\ }l@{}}
    0.075 \\ 0.075 \\ $0.000^{**}$ \end{tabular} & \begin{tabular}{@{\ }l@{}}
    $5.0 {\times} 10^5$ \\ $4.5 {\times} 10^{5}$  \\ $6.3 {\times} 10^5$ \end{tabular} & \begin{tabular}{@{\ }l@{}}
    3.05 \\ 2.83 \\ 3.68 \end{tabular} \\
    
      \textbf{This work} & $60$  & $5 {\times} 10^5$ & $2 \pi {\times} 1000$ & $1/12$ & $2.59 {\times} 10^{-4}$ & 2.6 & $\left\{\begin{tabular}{@{\ }l@{}}
    \emph{exp} \\ \emph{sim} \\ \emph{opt} \end{tabular}\right.$ & \begin{tabular}{@{\ }l@{}}
    8.5 \\ 8.5 \\ $8.4$ \end{tabular} & \begin{tabular}{@{\ }l@{}}
    0.22 \\ 0.22 \\ $0.18$ \end{tabular} & \begin{tabular}{@{\ }l@{}}
    $4.9 {\times} 10^4$ \\ $4.7 {\times} 10^4$  \\ $4.8 {\times} 10^4$ \end{tabular} & \begin{tabular}{@{\ }l@{}}
    3.95 \\ 3.85 \\ 3.93 \end{tabular} \\
  \hline
\end{tabular}
\caption{Experimental (\emph{exp}) and simulated (\emph{sim}) results from Refs.~\cite{Barrett_PRL_2001, Hung_PRA_2008} and this work. Also shown are the simulated optimal (\emph{opt}) results, which are found by optimizing $\eta$ and $\nu$ for the evaporation ramp, as in Fig.~\ref{fig:optimalgamma}. \\ $^{*}$\footnotesize{Value for $\eta$ is not provided in Ref.~\cite{Barrett_PRL_2001}; we use $\eta=7.2$ in simulation because it results in similar values for $N$, $T$, and total time of evaporation.}
\\$^{**}$\footnotesize{A result of optimal $\nu$ being equal to zero indicates a fixed frequency trap is ideal for evaporation. We do not consider values of $\nu<0$ (see footnote 27).}}
\label{tab:evapCompare}
\end{table*}
\endgroup

We acknowledge P. Wang and Q. Ma for the initial design and construction of our experimental apparatus and S. Dutta for helpful discussions. We are grateful for M. Olson's assistance with Fig.~2. We thank J.F. Clement, M. Zaiser, and M. Barrett for helpful communications, and D.S. Elliott for comments on this paper. The research was supported by DURIP-ARO Grant No. W911NF-08-1-0265 and the Miller Family Endowment. A.J.O. acknowledges support of the U.S. Deptartment of Defense NDSEG Fellowship Program.

\section*{Appendix}

\appendix
\section*{Loading the optical dipole trap}\label{sec:expDetails}

\setcounter{section}{1}

To load the optical dipole trap, we follow a procedure similar to that in Ref.~\cite{Clement_PRA_2009}, which was developed to allow for the loading of a dipole trap from a magneto-optical trap (MOT) when the dipole trap laser creates a strong ac Stark shift on the cooling transition of the MOT ($\approx -150$ MHz shift of the atomic transition relative to the cooling light frequency for our experiment). \footnote{This AC-Stark shift is due to the nearby $5P_{3/2}$ to $4D_{3/2}$ transition at 1529nm in $^{87}$Rb. The $5S_{1/2}$ to $5P_{3/2}$ transition is used as the cooling transition in the MOT.} To load the optical dipole trap, first load $\approx 1\times 10^8$ atoms in the MOT. During this time the cooling beams' detuning from the resonance, $\delta_{cooling}$, is set to -20 MHz with a total cooling beam power of 70 mW, the repump beam is set to resonance and an intensity $I_{repump}=1.8$ mW/cm$^2$, and the magnetic field gradient is set to $G_B=15$ G/cm. To maximize atoms loaded into the optical dipole trap, we apply a four stage compression and cooling process.  First, the $\delta_{cooling}$ is decreased in 10~ms to -30~MHz and held for 20~ms. Next, $G_B$ is increased to 51~G/cm in 5~ms and held for 25~ms. Then, $G_B$ is ramped to 1 G/cm in 5~ms. Simultaneously, $\delta_{cooling}$ is decreased to -35.5~MHz in 2~ms, and $I_{repump}$ is instantly decreased to a few $\mu$W/cm$^2$. Finally, an optical molasses stage is applied for 1 ms, where $I_{repump}$ is restored to 1.8~mW/cm$^2$, $G_B$ is ramped entirely off, and $\delta_{cooling}$ is ramped to -130.5~MHz. At the start of the 50 ms optical dipole trap loading stage, $\delta_{cooling}$ is changed to -195 MHz, $I_{repump}$ is dropped to a few $\mu$W/cm$^2$, and the dipole trap wide beam is turned on to full power (18~W). During that 50~ms, the dipole trap narrow beam is linearly ramped on to 9~W. After loading for 50~ms, the MOT cooling beams are turned off with the repump beams turned off a few hundreds of $\mu$s before the cooling to optically pump the atoms to the $F=1$ ground state. If only the wide beam is applied we observe $n =1\times 10^{11}$atoms/cm$^3$ and $T=10$ $\mu$K. For such initial conditions, the $\Gamma_{el}$ is too low for efficient evaporation. The addition of the narrow beam potential after loading increases $\Gamma_{el}$, aiding efficient evaporation \cite{StamperKurn_PRL_1998, Weber_Science_2003, ZYMa_JPB_2004}. After the 500 ms hold time in the optical dipole trap with the dimple on, there are $5\times 10^5$ atoms at 60 $\mu$K and a density of $3.5\times 10^{12}$ atoms/cm$^3$.

\bibliography{dipTrap}

\begin{thebibliography}{35}%
\makeatletter
\providecommand \@ifxundefined [1]{%
 \@ifx{#1\undefined}
}%
\providecommand \@ifnum [1]{%
 \ifnum #1\expandafter \@firstoftwo
 \else \expandafter \@secondoftwo
 \fi
}%
\providecommand \@ifx [1]{%
 \ifx #1\expandafter \@firstoftwo
 \else \expandafter \@secondoftwo
 \fi
}%
\providecommand \natexlab [1]{#1}%
\providecommand \enquote  [1]{``#1''}%
\providecommand \bibnamefont  [1]{#1}%
\providecommand \bibfnamefont [1]{#1}%
\providecommand \citenamefont [1]{#1}%
\providecommand \href@noop [0]{\@secondoftwo}%
\providecommand \href [0]{\begingroup \@sanitize@url \@href}%
\providecommand \@href[1]{\@@startlink{#1}\@@href}%
\providecommand \@@href[1]{\endgroup#1\@@endlink}%
\providecommand \@sanitize@url [0]{\catcode `\\12\catcode `\$12\catcode
  `\&12\catcode `\#12\catcode `\^12\catcode `\_12\catcode `\%12\relax}%
\providecommand \@@startlink[1]{}%
\providecommand \@@endlink[0]{}%
\providecommand \url  [0]{\begingroup\@sanitize@url \@url }%
\providecommand \@url [1]{\endgroup\@href {#1}{\urlprefix }}%
\providecommand \urlprefix  [0]{URL }%
\providecommand \Eprint [0]{\href }%
\providecommand \doibase [0]{http://dx.doi.org/}%
\providecommand \selectlanguage [0]{\@gobble}%
\providecommand \bibinfo  [0]{\@secondoftwo}%
\providecommand \bibfield  [0]{\@secondoftwo}%
\providecommand \translation [1]{[#1]}%
\providecommand \BibitemOpen [0]{}%
\providecommand \bibitemStop [0]{}%
\providecommand \bibitemNoStop [0]{.\EOS\space}%
\providecommand \EOS [0]{\spacefactor3000\relax}%
\providecommand \BibitemShut  [1]{\csname bibitem#1\endcsname}%
\let\auto@bib@innerbib\@empty
\bibitem [{\citenamefont {Chu}\ \emph {et~al.}(1986)\citenamefont {Chu},
  \citenamefont {Bjorkholm}, \citenamefont {Ashkin},\ and\ \citenamefont
  {Cable}}]{Chu_PRL_1986}%
  \BibitemOpen
  \bibfield  {author} {\bibinfo {author} {\bibfnamefont {S.}~\bibnamefont
  {Chu}}, \bibinfo {author} {\bibfnamefont {J.~E.}\ \bibnamefont {Bjorkholm}},
  \bibinfo {author} {\bibfnamefont {A.}~\bibnamefont {Ashkin}}, \ and\ \bibinfo
  {author} {\bibfnamefont {A.}~\bibnamefont {Cable}},\ }\href {\doibase
  10.1103/PhysRevLett.57.314} {\bibfield  {journal} {\bibinfo  {journal} {Phys.
  Rev. Lett.}\ }\textbf {\bibinfo {volume} {57}},\ \bibinfo {pages} {314}
  (\bibinfo {year} {1986})}\BibitemShut {NoStop}%
\bibitem [{\citenamefont {Lovelace}\ \emph {et~al.}(1985)\citenamefont
  {Lovelace}, \citenamefont {Mehanian}, \citenamefont {Tommila},\ and\
  \citenamefont {Lee}}]{Lovelace_Nature_1985}%
  \BibitemOpen
  \bibfield  {author} {\bibinfo {author} {\bibfnamefont {R.}~\bibnamefont
  {Lovelace}}, \bibinfo {author} {\bibfnamefont {C.}~\bibnamefont {Mehanian}},
  \bibinfo {author} {\bibfnamefont {T.}~\bibnamefont {Tommila}}, \ and\
  \bibinfo {author} {\bibfnamefont {D.}~\bibnamefont {Lee}},\ }\href@noop {}
  {\bibfield  {journal} {\bibinfo  {journal} {Nature}\ }\textbf {\bibinfo
  {volume} {318}},\ \bibinfo {pages} {30} (\bibinfo {year} {1985})}\BibitemShut
  {NoStop}%
\bibitem [{\citenamefont {Masuhara}\ \emph {et~al.}(1988)\citenamefont
  {Masuhara}, \citenamefont {Doyle}, \citenamefont {Sandberg}, \citenamefont
  {Kleppner}, \citenamefont {Greytak}, \citenamefont {Hess},\ and\
  \citenamefont {Kochanski}}]{Masuhara_PRL_1988}%
  \BibitemOpen
  \bibfield  {author} {\bibinfo {author} {\bibfnamefont {N.}~\bibnamefont
  {Masuhara}}, \bibinfo {author} {\bibfnamefont {J.~M.}\ \bibnamefont {Doyle}},
  \bibinfo {author} {\bibfnamefont {J.~C.}\ \bibnamefont {Sandberg}}, \bibinfo
  {author} {\bibfnamefont {D.}~\bibnamefont {Kleppner}}, \bibinfo {author}
  {\bibfnamefont {T.~J.}\ \bibnamefont {Greytak}}, \bibinfo {author}
  {\bibfnamefont {H.~F.}\ \bibnamefont {Hess}}, \ and\ \bibinfo {author}
  {\bibfnamefont {G.~P.}\ \bibnamefont {Kochanski}},\ }\href {\doibase
  10.1103/PhysRevLett.61.935} {\bibfield  {journal} {\bibinfo  {journal} {Phys.
  Rev. Lett.}\ }\textbf {\bibinfo {volume} {61}},\ \bibinfo {pages} {935}
  (\bibinfo {year} {1988})}\BibitemShut {NoStop}%
\bibitem [{\citenamefont {Anderson}\ \emph {et~al.}(1995)\citenamefont
  {Anderson}, \citenamefont {Ensher}, \citenamefont {Matthews}, \citenamefont
  {Wieman},\ and\ \citenamefont {Cornell}}]{Anderson_Science_1995}%
  \BibitemOpen
  \bibfield  {author} {\bibinfo {author} {\bibfnamefont {M.~H.}\ \bibnamefont
  {Anderson}}, \bibinfo {author} {\bibfnamefont {J.~R.}\ \bibnamefont
  {Ensher}}, \bibinfo {author} {\bibfnamefont {M.~R.}\ \bibnamefont
  {Matthews}}, \bibinfo {author} {\bibfnamefont {C.~E.}\ \bibnamefont
  {Wieman}}, \ and\ \bibinfo {author} {\bibfnamefont {E.~A.}\ \bibnamefont
  {Cornell}},\ }\href@noop {} {\bibfield  {journal} {\bibinfo  {journal}
  {Science}\ }\textbf {\bibinfo {volume} {269}},\ \bibinfo {pages} {198}
  (\bibinfo {year} {1995})}\BibitemShut {NoStop}%
\bibitem [{\citenamefont {Davis}\ \emph
  {et~al.}(1995{\natexlab{a}})\citenamefont {Davis}, \citenamefont {Mewes},
  \citenamefont {Andrews}, \citenamefont {van Druten}, \citenamefont {Durfee},
  \citenamefont {Kurn},\ and\ \citenamefont {Ketterle}}]{Davis_PRL_1995}%
  \BibitemOpen
  \bibfield  {author} {\bibinfo {author} {\bibfnamefont {K.~B.}\ \bibnamefont
  {Davis}}, \bibinfo {author} {\bibfnamefont {M.~O.}\ \bibnamefont {Mewes}},
  \bibinfo {author} {\bibfnamefont {M.~R.}\ \bibnamefont {Andrews}}, \bibinfo
  {author} {\bibfnamefont {N.~J.}\ \bibnamefont {van Druten}}, \bibinfo
  {author} {\bibfnamefont {D.~S.}\ \bibnamefont {Durfee}}, \bibinfo {author}
  {\bibfnamefont {D.~M.}\ \bibnamefont {Kurn}}, \ and\ \bibinfo {author}
  {\bibfnamefont {W.}~\bibnamefont {Ketterle}},\ }\href {\doibase
  10.1103/PhysRevLett.75.3969} {\bibfield  {journal} {\bibinfo  {journal}
  {Phys. Rev. Lett.}\ }\textbf {\bibinfo {volume} {75}},\ \bibinfo {pages}
  {3969} (\bibinfo {year} {1995}{\natexlab{a}})}\BibitemShut {NoStop}%
\bibitem [{\citenamefont {Bradley}\ \emph {et~al.}(1995)\citenamefont
  {Bradley}, \citenamefont {Sackett}, \citenamefont {Tollett},\ and\
  \citenamefont {Hulet}}]{Bradley_PRL_1995}%
  \BibitemOpen
  \bibfield  {author} {\bibinfo {author} {\bibfnamefont {C.~C.}\ \bibnamefont
  {Bradley}}, \bibinfo {author} {\bibfnamefont {C.~A.}\ \bibnamefont
  {Sackett}}, \bibinfo {author} {\bibfnamefont {J.~J.}\ \bibnamefont
  {Tollett}}, \ and\ \bibinfo {author} {\bibfnamefont {R.~G.}\ \bibnamefont
  {Hulet}},\ }\href {\doibase 10.1103/PhysRevLett.75.1687} {\bibfield
  {journal} {\bibinfo  {journal} {Phys. Rev. Lett.}\ }\textbf {\bibinfo
  {volume} {75}},\ \bibinfo {pages} {1687} (\bibinfo {year}
  {1995})}\BibitemShut {NoStop}%
\bibitem [{\citenamefont {Bradley}\ \emph {et~al.}(1997)\citenamefont
  {Bradley}, \citenamefont {Sackett},\ and\ \citenamefont
  {Hulet}}]{Bradley_PRL_1997}%
  \BibitemOpen
  \bibfield  {author} {\bibinfo {author} {\bibfnamefont {C.~C.}\ \bibnamefont
  {Bradley}}, \bibinfo {author} {\bibfnamefont {C.~A.}\ \bibnamefont
  {Sackett}}, \ and\ \bibinfo {author} {\bibfnamefont {R.~G.}\ \bibnamefont
  {Hulet}},\ }\href {\doibase 10.1103/PhysRevLett.78.985} {\bibfield  {journal}
  {\bibinfo  {journal} {Phys. Rev. Lett.}\ }\textbf {\bibinfo {volume} {78}},\
  \bibinfo {pages} {985} (\bibinfo {year} {1997})}\BibitemShut {NoStop}%
\bibitem [{\citenamefont {DeMarco}\ and\ \citenamefont
  {Jin}(1999)}]{DeMarco_Science_1999}%
  \BibitemOpen
  \bibfield  {author} {\bibinfo {author} {\bibfnamefont {B.}~\bibnamefont
  {DeMarco}}\ and\ \bibinfo {author} {\bibfnamefont {D.~S.}\ \bibnamefont
  {Jin}},\ }\href@noop {} {\bibfield  {journal} {\bibinfo  {journal} {Science}\
  }\textbf {\bibinfo {volume} {285}},\ \bibinfo {pages} {1703} (\bibinfo {year}
  {1999})}\BibitemShut {NoStop}%
\bibitem [{\citenamefont {Ketterle}\ and\ \citenamefont {van
  Druten}(1996)}]{Ketterle_AdvAMO_1996}%
  \BibitemOpen
  \bibfield  {author} {\bibinfo {author} {\bibfnamefont {W.}~\bibnamefont
  {Ketterle}}\ and\ \bibinfo {author} {\bibfnamefont {N.~J.}\ \bibnamefont {van
  Druten}}\ }(\bibinfo  {publisher} {Academic Press},\ \bibinfo {year} {1996})\
  pp.\ \bibinfo {pages} {181 -- 236}\BibitemShut {NoStop}%
\bibitem [{\citenamefont {Luiten}\ \emph {et~al.}(1996)\citenamefont {Luiten},
  \citenamefont {Reynolds},\ and\ \citenamefont {Walraven}}]{Luiten_PRA_1996}%
  \BibitemOpen
  \bibfield  {author} {\bibinfo {author} {\bibfnamefont {O.~J.}\ \bibnamefont
  {Luiten}}, \bibinfo {author} {\bibfnamefont {M.~W.}\ \bibnamefont
  {Reynolds}}, \ and\ \bibinfo {author} {\bibfnamefont {J.~T.~M.}\ \bibnamefont
  {Walraven}},\ }\href {\doibase 10.1103/PhysRevA.53.381} {\bibfield  {journal}
  {\bibinfo  {journal} {Phys. Rev. A}\ }\textbf {\bibinfo {volume} {53}},\
  \bibinfo {pages} {381} (\bibinfo {year} {1996})}\BibitemShut {NoStop}%
\bibitem [{\citenamefont {Surkov}\ \emph {et~al.}(1996)\citenamefont {Surkov},
  \citenamefont {Walraven},\ and\ \citenamefont
  {Shlyapnikov}}]{Surkov_PRA_1996}%
  \BibitemOpen
  \bibfield  {author} {\bibinfo {author} {\bibfnamefont {E.~L.}\ \bibnamefont
  {Surkov}}, \bibinfo {author} {\bibfnamefont {J.~T.~M.}\ \bibnamefont
  {Walraven}}, \ and\ \bibinfo {author} {\bibfnamefont {G.~V.}\ \bibnamefont
  {Shlyapnikov}},\ }\href {\doibase 10.1103/PhysRevA.53.3403} {\bibfield
  {journal} {\bibinfo  {journal} {Phys. Rev. A}\ }\textbf {\bibinfo {volume}
  {53}},\ \bibinfo {pages} {3403} (\bibinfo {year} {1996})}\BibitemShut
  {NoStop}%
\bibitem [{\citenamefont {Walraven}(1996)}]{Walraven_book_1996}%
  \BibitemOpen
  \bibfield  {author} {\bibinfo {author} {\bibfnamefont {J.~T.~M.}\
  \bibnamefont {Walraven}},\ }in\ \href@noop {} {\emph {\bibinfo {booktitle}
  {Quantum dynamics of simple systems}}},\ \bibinfo {editor} {edited by\
  \bibinfo {editor} {\bibfnamefont {G.-L.}\ \bibnamefont {Oppo}}, \bibinfo
  {editor} {\bibfnamefont {S.}~\bibnamefont {Barnett}}, \bibinfo {editor}
  {\bibfnamefont {E.}~\bibnamefont {Riis}}, \ and\ \bibinfo {editor}
  {\bibfnamefont {M.}~\bibnamefont {Wilkinson}}}\ (\bibinfo  {publisher} {IOP:
  Bristol},\ \bibinfo {year} {1996})\ pp.\ \bibinfo {pages}
  {315--352}\BibitemShut {NoStop}%
\bibitem [{\citenamefont {Sackett}\ \emph {et~al.}(1997)\citenamefont
  {Sackett}, \citenamefont {Bradley},\ and\ \citenamefont
  {Hulet}}]{Sackett_PRA_1997}%
  \BibitemOpen
  \bibfield  {author} {\bibinfo {author} {\bibfnamefont {C.~A.}\ \bibnamefont
  {Sackett}}, \bibinfo {author} {\bibfnamefont {C.~C.}\ \bibnamefont
  {Bradley}}, \ and\ \bibinfo {author} {\bibfnamefont {R.~G.}\ \bibnamefont
  {Hulet}},\ }\href {\doibase 10.1103/PhysRevA.55.3797} {\bibfield  {journal}
  {\bibinfo  {journal} {Phys. Rev. A}\ }\textbf {\bibinfo {volume} {55}},\
  \bibinfo {pages} {3797} (\bibinfo {year} {1997})}\BibitemShut {NoStop}%
\bibitem [{\citenamefont {Yamashita}\ \emph {et~al.}(2003)\citenamefont
  {Yamashita}, \citenamefont {Koashi}, \citenamefont {Mukai}, \citenamefont
  {Mitsunaga}, \citenamefont {Imoto},\ and\ \citenamefont
  {Mukai}}]{Yamashita_PRA_2003}%
  \BibitemOpen
  \bibfield  {author} {\bibinfo {author} {\bibfnamefont {M.}~\bibnamefont
  {Yamashita}}, \bibinfo {author} {\bibfnamefont {M.}~\bibnamefont {Koashi}},
  \bibinfo {author} {\bibfnamefont {T.}~\bibnamefont {Mukai}}, \bibinfo
  {author} {\bibfnamefont {M.}~\bibnamefont {Mitsunaga}}, \bibinfo {author}
  {\bibfnamefont {N.}~\bibnamefont {Imoto}}, \ and\ \bibinfo {author}
  {\bibfnamefont {T.}~\bibnamefont {Mukai}},\ }\href {\doibase
  10.1103/PhysRevA.67.023601} {\bibfield  {journal} {\bibinfo  {journal} {Phys.
  Rev. A}\ }\textbf {\bibinfo {volume} {67}},\ \bibinfo {pages} {023601}
  (\bibinfo {year} {2003})}\BibitemShut {NoStop}%
\bibitem [{\citenamefont {Davis}\ \emph
  {et~al.}(1995{\natexlab{b}})\citenamefont {Davis}, \citenamefont {Mewes},
  \citenamefont {Joffe}, \citenamefont {Andrews},\ and\ \citenamefont
  {Ketterle}}]{Davis_PRL_1995a}%
  \BibitemOpen
  \bibfield  {author} {\bibinfo {author} {\bibfnamefont {K.~B.}\ \bibnamefont
  {Davis}}, \bibinfo {author} {\bibfnamefont {M.-O.}\ \bibnamefont {Mewes}},
  \bibinfo {author} {\bibfnamefont {M.~A.}\ \bibnamefont {Joffe}}, \bibinfo
  {author} {\bibfnamefont {M.~R.}\ \bibnamefont {Andrews}}, \ and\ \bibinfo
  {author} {\bibfnamefont {W.}~\bibnamefont {Ketterle}},\ }\href {\doibase
  10.1103/PhysRevLett.74.5202} {\bibfield  {journal} {\bibinfo  {journal}
  {Phys. Rev. Lett.}\ }\textbf {\bibinfo {volume} {74}},\ \bibinfo {pages}
  {5202} (\bibinfo {year} {1995}{\natexlab{b}})}\BibitemShut {NoStop}%
\bibitem [{\citenamefont {Barrett}\ \emph {et~al.}(2001)\citenamefont
  {Barrett}, \citenamefont {Sauer},\ and\ \citenamefont
  {Chapman}}]{Barrett_PRL_2001}%
  \BibitemOpen
  \bibfield  {author} {\bibinfo {author} {\bibfnamefont {M.~D.}\ \bibnamefont
  {Barrett}}, \bibinfo {author} {\bibfnamefont {J.~A.}\ \bibnamefont {Sauer}},
  \ and\ \bibinfo {author} {\bibfnamefont {M.~S.}\ \bibnamefont {Chapman}},\
  }\href {\doibase 10.1103/PhysRevLett.87.010404} {\bibfield  {journal}
  {\bibinfo  {journal} {Phys. Rev. Lett.}\ }\textbf {\bibinfo {volume} {87}},\
  \bibinfo {pages} {010404} (\bibinfo {year} {2001})}\BibitemShut {NoStop}%
\bibitem [{\citenamefont {O'Hara}\ \emph {et~al.}(2001)\citenamefont {O'Hara},
  \citenamefont {Gehm}, \citenamefont {Granade},\ and\ \citenamefont
  {Thomas}}]{OHara_PRA_2001}%
  \BibitemOpen
  \bibfield  {author} {\bibinfo {author} {\bibfnamefont {K.~M.}\ \bibnamefont
  {O'Hara}}, \bibinfo {author} {\bibfnamefont {M.~E.}\ \bibnamefont {Gehm}},
  \bibinfo {author} {\bibfnamefont {S.~R.}\ \bibnamefont {Granade}}, \ and\
  \bibinfo {author} {\bibfnamefont {J.~E.}\ \bibnamefont {Thomas}},\ }\href
  {\doibase 10.1103/PhysRevA.64.051403} {\bibfield  {journal} {\bibinfo
  {journal} {Phys. Rev. A}\ }\textbf {\bibinfo {volume} {64}},\ \bibinfo
  {pages} {051403} (\bibinfo {year} {2001})}\BibitemShut {NoStop}%
\bibitem [{\citenamefont {Kinoshita}\ \emph {et~al.}(2005)\citenamefont
  {Kinoshita}, \citenamefont {Wenger},\ and\ \citenamefont
  {Weiss}}]{Kinoshita_PRA_2005}%
  \BibitemOpen
  \bibfield  {author} {\bibinfo {author} {\bibfnamefont {T.}~\bibnamefont
  {Kinoshita}}, \bibinfo {author} {\bibfnamefont {T.}~\bibnamefont {Wenger}}, \
  and\ \bibinfo {author} {\bibfnamefont {D.~S.}\ \bibnamefont {Weiss}},\ }\href
  {\doibase 10.1103/PhysRevA.71.011602} {\bibfield  {journal} {\bibinfo
  {journal} {Phys. Rev. A}\ }\textbf {\bibinfo {volume} {71}},\ \bibinfo
  {pages} {011602} (\bibinfo {year} {2005})}\BibitemShut {NoStop}%
\bibitem [{\citenamefont {Hung}\ \emph {et~al.}(2008)\citenamefont {Hung},
  \citenamefont {Zhang}, \citenamefont {Gemelke},\ and\ \citenamefont
  {Chin}}]{Hung_PRA_2008}%
  \BibitemOpen
  \bibfield  {author} {\bibinfo {author} {\bibfnamefont {C.-L.}\ \bibnamefont
  {Hung}}, \bibinfo {author} {\bibfnamefont {X.}~\bibnamefont {Zhang}},
  \bibinfo {author} {\bibfnamefont {N.}~\bibnamefont {Gemelke}}, \ and\
  \bibinfo {author} {\bibfnamefont {C.}~\bibnamefont {Chin}},\ }\href {\doibase
  10.1103/PhysRevA.78.011604} {\bibfield  {journal} {\bibinfo  {journal} {Phys.
  Rev. A}\ }\textbf {\bibinfo {volume} {78}},\ \bibinfo {pages} {011604}
  (\bibinfo {year} {2008})}\BibitemShut {NoStop}%
\bibitem [{\citenamefont {Cl\'ement}\ \emph {et~al.}(2009)\citenamefont
  {Cl\'ement}, \citenamefont {Brantut}, \citenamefont {Robert-de
  Saint-Vincent}, \citenamefont {Nyman}, \citenamefont {Aspect}, \citenamefont
  {Bourdel},\ and\ \citenamefont {Bouyer}}]{Clement_PRA_2009}%
  \BibitemOpen
  \bibfield  {author} {\bibinfo {author} {\bibfnamefont {J.-F.}\ \bibnamefont
  {Cl\'ement}}, \bibinfo {author} {\bibfnamefont {J.-P.}\ \bibnamefont
  {Brantut}}, \bibinfo {author} {\bibfnamefont {M.}~\bibnamefont {Robert-de
  Saint-Vincent}}, \bibinfo {author} {\bibfnamefont {R.~A.}\ \bibnamefont
  {Nyman}}, \bibinfo {author} {\bibfnamefont {A.}~\bibnamefont {Aspect}},
  \bibinfo {author} {\bibfnamefont {T.}~\bibnamefont {Bourdel}}, \ and\
  \bibinfo {author} {\bibfnamefont {P.}~\bibnamefont {Bouyer}},\ }\href
  {\doibase 10.1103/PhysRevA.79.061406} {\bibfield  {journal} {\bibinfo
  {journal} {Phys. Rev. A}\ }\textbf {\bibinfo {volume} {79}},\ \bibinfo
  {pages} {061406} (\bibinfo {year} {2009})}\BibitemShut {NoStop}%
\bibitem [{\citenamefont {Arnold}\ and\ \citenamefont
  {Barrett}(2011)}]{Arnold_OptComm_2011}%
  \BibitemOpen
  \bibfield  {author} {\bibinfo {author} {\bibfnamefont {K.}~\bibnamefont
  {Arnold}}\ and\ \bibinfo {author} {\bibfnamefont {M.}~\bibnamefont
  {Barrett}},\ }\href@noop {} {\bibfield  {journal} {\bibinfo  {journal}
  {Optics Communications}\ }\textbf {\bibinfo {volume} {284}},\ \bibinfo
  {pages} {3288} (\bibinfo {year} {2011})}\BibitemShut {NoStop}%
\bibitem [{\citenamefont {Pinkse}\ \emph {et~al.}(1998)\citenamefont {Pinkse},
  \citenamefont {Mosk}, \citenamefont {Weidem\"uller}, \citenamefont
  {Reynolds}, \citenamefont {Hijmans},\ and\ \citenamefont
  {Walraven}}]{Pinkse_PRA_1998}%
  \BibitemOpen
  \bibfield  {author} {\bibinfo {author} {\bibfnamefont {P.~W.~H.}\
  \bibnamefont {Pinkse}}, \bibinfo {author} {\bibfnamefont {A.}~\bibnamefont
  {Mosk}}, \bibinfo {author} {\bibfnamefont {M.}~\bibnamefont {Weidem\"uller}},
  \bibinfo {author} {\bibfnamefont {M.~W.}\ \bibnamefont {Reynolds}}, \bibinfo
  {author} {\bibfnamefont {T.~W.}\ \bibnamefont {Hijmans}}, \ and\ \bibinfo
  {author} {\bibfnamefont {J.~T.~M.}\ \bibnamefont {Walraven}},\ }\href
  {\doibase 10.1103/PhysRevA.57.4747} {\bibfield  {journal} {\bibinfo
  {journal} {Phys. Rev. A}\ }\textbf {\bibinfo {volume} {57}},\ \bibinfo
  {pages} {4747} (\bibinfo {year} {1998})}\BibitemShut {NoStop}%
\bibitem [{\citenamefont {Yamashita}(2001)}]{Yamashita_LP_2003}%
  \BibitemOpen
  \bibfield  {author} {\bibinfo {author} {\bibfnamefont {M.}~\bibnamefont
  {Yamashita}},\ }\href@noop {} {\bibfield  {journal} {\bibinfo  {journal}
  {Laser Physics}\ }\textbf {\bibinfo {volume} {14}},\ \bibinfo {pages} {597}
  (\bibinfo {year} {2001})}\BibitemShut {NoStop}%
\bibitem [{\citenamefont {van Kempen}\ \emph {et~al.}(2002)\citenamefont {van
  Kempen}, \citenamefont {Kokkelmans}, \citenamefont {Heinzen},\ and\
  \citenamefont {Verhaar}}]{VanKempen_PRL_2002}%
  \BibitemOpen
  \bibfield  {author} {\bibinfo {author} {\bibfnamefont {E.~G.~M.}\
  \bibnamefont {van Kempen}}, \bibinfo {author} {\bibfnamefont {S.~J. J.
  M.~F.}\ \bibnamefont {Kokkelmans}}, \bibinfo {author} {\bibfnamefont {D.~J.}\
  \bibnamefont {Heinzen}}, \ and\ \bibinfo {author} {\bibfnamefont {B.~J.}\
  \bibnamefont {Verhaar}},\ }\href {\doibase 10.1103/PhysRevLett.88.093201}
  {\bibfield  {journal} {\bibinfo  {journal} {Phys. Rev. Lett.}\ }\textbf
  {\bibinfo {volume} {88}},\ \bibinfo {pages} {093201} (\bibinfo {year}
  {2002})}\BibitemShut {NoStop}%
\bibitem [{\citenamefont {Burt}\ \emph {et~al.}(1997)\citenamefont {Burt},
  \citenamefont {Ghrist}, \citenamefont {Myatt}, \citenamefont {Holland},
  \citenamefont {Cornell},\ and\ \citenamefont {Wieman}}]{Burt_PRL_1997}%
  \BibitemOpen
  \bibfield  {author} {\bibinfo {author} {\bibfnamefont {E.~A.}\ \bibnamefont
  {Burt}}, \bibinfo {author} {\bibfnamefont {R.~W.}\ \bibnamefont {Ghrist}},
  \bibinfo {author} {\bibfnamefont {C.~J.}\ \bibnamefont {Myatt}}, \bibinfo
  {author} {\bibfnamefont {M.~J.}\ \bibnamefont {Holland}}, \bibinfo {author}
  {\bibfnamefont {E.~A.}\ \bibnamefont {Cornell}}, \ and\ \bibinfo {author}
  {\bibfnamefont {C.~E.}\ \bibnamefont {Wieman}},\ }\href {\doibase
  10.1103/PhysRevLett.79.337} {\bibfield  {journal} {\bibinfo  {journal} {Phys.
  Rev. Lett.}\ }\textbf {\bibinfo {volume} {79}},\ \bibinfo {pages} {337}
  (\bibinfo {year} {1997})}\BibitemShut {NoStop}%
\bibitem [{Note1()}]{Note1}%
  \BibitemOpen
  \bibinfo {note} {The results show the importance of finding the trap averaged
  values given the dimensionality and trap type used in the experiment to
  obtain quantitative agreement between the theory and experiments. $\Gamma
  _{el}$ is sometimes reported as $n \sigma \protect \mathaccentV {bar}016{v}$,
  but in a 3D harmonic potential is reduced by $2\protect \sqrt {2}$.
  Similarly, since three-body collisions occur in the denser, less energetic
  regions of the trap, the average energy loss per atom from three-body
  collisions is two-thirds of the average energy of an atom in the
  trap.}\BibitemShut {Stop}%
\bibitem [{Note2()}]{Note2}%
  \BibitemOpen
  \bibinfo {note} {For the initial conditions of our experiment, we find that
  implementing a variable $\nu (t)$ during evaporation gives only a small gain
  ($\approx 0.1$) in the numerically computed $\gamma _{eff}$.}\BibitemShut
  {Stop}%
\bibitem [{Note3()}]{Note3}%
  \BibitemOpen
  \bibinfo {note} {Values of $\nu < 0$ are theoretically possible to treat, but
  are not considered here as experimental limitations in available laser power
  usually restrict $\nu $ to greater than $0$.}\BibitemShut {Stop}%
\bibitem [{\citenamefont {Miesner}\ \emph {et~al.}(1998)\citenamefont
  {Miesner}, \citenamefont {Stamper-Kurn}, \citenamefont {Andrews},
  \citenamefont {Durfee}, \citenamefont {Inouye},\ and\ \citenamefont
  {Ketterle}}]{Miesner_Science_1998}%
  \BibitemOpen
  \bibfield  {author} {\bibinfo {author} {\bibfnamefont {H.~J.}\ \bibnamefont
  {Miesner}}, \bibinfo {author} {\bibfnamefont {D.~M.}\ \bibnamefont
  {Stamper-Kurn}}, \bibinfo {author} {\bibfnamefont {M.~R.}\ \bibnamefont
  {Andrews}}, \bibinfo {author} {\bibfnamefont {D.~S.}\ \bibnamefont {Durfee}},
  \bibinfo {author} {\bibfnamefont {S.}~\bibnamefont {Inouye}}, \ and\ \bibinfo
  {author} {\bibfnamefont {W.}~\bibnamefont {Ketterle}},\ }\href@noop {}
  {\bibfield  {journal} {\bibinfo  {journal} {Science}\ }\textbf {\bibinfo
  {volume} {279}},\ \bibinfo {pages} {1005} (\bibinfo {year}
  {1998})}\BibitemShut {NoStop}%
\bibitem [{\citenamefont {K\"ohl}\ \emph {et~al.}(2002)\citenamefont {K\"ohl},
  \citenamefont {Davis}, \citenamefont {Gardiner}, \citenamefont {H\"ansch},\
  and\ \citenamefont {Esslinger}}]{Kohl_PRL_2002}%
  \BibitemOpen
  \bibfield  {author} {\bibinfo {author} {\bibfnamefont {M.}~\bibnamefont
  {K\"ohl}}, \bibinfo {author} {\bibfnamefont {M.~J.}\ \bibnamefont {Davis}},
  \bibinfo {author} {\bibfnamefont {C.~W.}\ \bibnamefont {Gardiner}}, \bibinfo
  {author} {\bibfnamefont {T.~W.}\ \bibnamefont {H\"ansch}}, \ and\ \bibinfo
  {author} {\bibfnamefont {T.}~\bibnamefont {Esslinger}},\ }\href {\doibase
  10.1103/PhysRevLett.88.080402} {\bibfield  {journal} {\bibinfo  {journal}
  {Phys. Rev. Lett.}\ }\textbf {\bibinfo {volume} {88}},\ \bibinfo {pages}
  {080402} (\bibinfo {year} {2002})}\BibitemShut {NoStop}%
\bibitem [{Note4()}]{Note4}%
  \BibitemOpen
  \bibinfo {note} {The total time of evaporation is important when considering
  the cycle time of an apparatus which produces BECs. The optimized schemes
  presented here take $0.7$ and $10$ seconds for the Barrett \protect \emph {et
  al.} and Hung \protect \emph {et al.} experiments, respectively.}\BibitemShut
  {Stop}%
\bibitem [{Note5()}]{Note5}%
  \BibitemOpen
  \bibinfo {note} {This AC-Stark shift is due to the nearby $5P_{3/2}$ to
  $4D_{3/2}$ transition at 1529nm in $^{87}$Rb. The $5S_{1/2}$ to $5P_{3/2}$
  transition is used as the cooling transition in the MOT.}\BibitemShut {Stop}%
\bibitem [{\citenamefont {Stamper-Kurn}\ \emph {et~al.}(1998)\citenamefont
  {Stamper-Kurn}, \citenamefont {Miesner}, \citenamefont {Chikkatur},
  \citenamefont {Inouye}, \citenamefont {Stenger},\ and\ \citenamefont
  {Ketterle}}]{StamperKurn_PRL_1998}%
  \BibitemOpen
  \bibfield  {author} {\bibinfo {author} {\bibfnamefont {D.~M.}\ \bibnamefont
  {Stamper-Kurn}}, \bibinfo {author} {\bibfnamefont {H.-J.}\ \bibnamefont
  {Miesner}}, \bibinfo {author} {\bibfnamefont {A.~P.}\ \bibnamefont
  {Chikkatur}}, \bibinfo {author} {\bibfnamefont {S.}~\bibnamefont {Inouye}},
  \bibinfo {author} {\bibfnamefont {J.}~\bibnamefont {Stenger}}, \ and\
  \bibinfo {author} {\bibfnamefont {W.}~\bibnamefont {Ketterle}},\ }\href
  {\doibase 10.1103/PhysRevLett.81.2194} {\bibfield  {journal} {\bibinfo
  {journal} {Phys. Rev. Lett.}\ }\textbf {\bibinfo {volume} {81}},\ \bibinfo
  {pages} {2194} (\bibinfo {year} {1998})}\BibitemShut {NoStop}%
\bibitem [{\citenamefont {Weber}\ \emph {et~al.}(2003)\citenamefont {Weber},
  \citenamefont {Herbig}, \citenamefont {Mark}, \citenamefont {N\"{a}gerl},\
  and\ \citenamefont {Grimm}}]{Weber_Science_2003}%
  \BibitemOpen
  \bibfield  {author} {\bibinfo {author} {\bibfnamefont {T.}~\bibnamefont
  {Weber}}, \bibinfo {author} {\bibfnamefont {J.}~\bibnamefont {Herbig}},
  \bibinfo {author} {\bibfnamefont {M.}~\bibnamefont {Mark}}, \bibinfo {author}
  {\bibfnamefont {H.-C.}\ \bibnamefont {N\"{a}gerl}}, \ and\ \bibinfo {author}
  {\bibfnamefont {R.}~\bibnamefont {Grimm}},\ }\href@noop {} {\bibfield
  {journal} {\bibinfo  {journal} {Science}\ }\textbf {\bibinfo {volume}
  {299}},\ \bibinfo {pages} {232} (\bibinfo {year} {2003})}\BibitemShut
  {NoStop}%
\bibitem [{\citenamefont {Ma}\ \emph {et~al.}(2004)\citenamefont {Ma},
  \citenamefont {Foot},\ and\ \citenamefont {Cornish}}]{ZYMa_JPB_2004}%
  \BibitemOpen
  \bibfield  {author} {\bibinfo {author} {\bibfnamefont {Z.}~\bibnamefont
  {Ma}}, \bibinfo {author} {\bibfnamefont {C.}~\bibnamefont {Foot}}, \ and\
  \bibinfo {author} {\bibfnamefont {S.}~\bibnamefont {Cornish}},\ }\href@noop
  {} {\bibfield  {journal} {\bibinfo  {journal} {Journal of Physics B}\
  }\textbf {\bibinfo {volume} {37}},\ \bibinfo {pages} {3187} (\bibinfo {year}
  {2004})}\BibitemShut {NoStop}%
\end{thebibliography}%
\end{document}